\documentclass[]{spie}
\usepackage[dvips]{color}
\usepackage{aastex_hack}
\usepackage{amssymb}
\usepackage{epsfig}
\usepackage[latin1]{inputenc}
\usepackage{graphicx}
\usepackage{times}
\usepackage{amssymb}
\usepackage{color}
\usepackage[figuresright]{rotating}
\usepackage{amsmath}
\usepackage{url}

\newcommand{\figurewidth} {0.48\textwidth}
\newcommand{\Deg}{${}^{\circ}$}

\title{CCCP: A CCD Controller for Counting Photons}
\author{Olivier Daigle\supit{a,b,c}, Jean-Luc Gach\supit{b}, Christian Guillaume\supit{d}, Simon Lessard\supit{c}, Claude Carignan\supit{a,b,e}, Sébastien Blais-Ouellette\supit{c}
\skiplinehalf
\supit{a} Laboratoire d'Astrophysique Exp\'erimentale, D\'epartement de physique, Universit\'e de Montr\'eal, C.P. 6128 Succ. Centre-Ville, Montr\'eal, QC, Canada, H3C 3J7;\\
\supit{b} Laboratoire d'Astrophysique de Marseille, Observatoire Astronomique de Marseille-Provence, Technopôle de Château-Gombert,  38, rue Frédéric Joliot-Curie, 13388 Marseille, France;\\
\supit{c} Photon etc., 5155 Decelles Avenue, Pavillon J.A Bombardier, Montr\'eal, Qu\'ebec, Canada, H3T 2B1;\\
\supit{d} Observatoire de Haute-Provence, 04870 St-Michel l'observatoire, France;\\
\supit{e} Observatoire d'Astrophysique de l'Université de Ouagadougou, BP 7021, Ouagadougou 03, Burkina Faso.}

\authorinfo{Send correspondance to O.D.: odaigle@astro.umontreal.ca}

\begin{document}
\maketitle

\begin{abstract}
CCCP, a CCD Controller for Counting Photons, is presented. This new
controller uses a totally new clocking architecture and allows to drive the
CCD in a novel way. Its design is optimized for the driving of EMCCDs at up to 20MHz of pixel rate and fast vertical transfer. Using this controller, the dominant source of noise of EMCCDs at low flux level and high frame rate, the Clock Induced Charges, were reduced to 0.001 -- 0.0018 electron/pixel/frame (depending of the electron multiplying gain), making efficient photon counting possible. CCCP will be deployed in 2009 on the ESO NTT through the 3D-NTT\cite{2007spts.conf...15M} project and on the SOAR through the BTFI project.\\
\end{abstract}


\section{INTRODUCTION}
\label{sect:intro}
Electron Multiplying Charge Coupled Devices (EMCCD) allows one to apply a gain to the pixel's charge before it reaches the noisy output amplifier where the charge-tension conversion is made \cite{techreport-minimal2}. A gain $G$ in the charge domain affects the effective readout noise by the relation $\sigma_{eff} = \frac{\sigma_{real}}{G}$. Sub-electron effective readout noise levels are thus achievable. However, the electron multiplying process is stochastic. This statistical behaviour adds an excess noise factor 
that reaches a value of $2^{1/2}$ at high gains \cite{stanford}. The effect on the signal-to-noise ratio (SNR) of the system is the same as if the quantum efficiency (QE) of the CCD would be halved.

Some authors proposed offline data processing to lower the excess noise factor induced by the multiplication register\cite{2008MNRAS.386.2262L, 2003MNRAS.345..985B}. However, the only way to overcome the excess noise factor without any \textit{a priori} knowledge or stability assumption on the signal is to consider the pixel binary by applying a single threshold to the output signal. This way, only one photon per pixel per frame can be counted and the full QE of the silicon can be recovered, making the EMCCD a theoretically perfect photon counting device. The highest flux rate that can be observed in this mode will thus depend of the frame rate at which the EMCCD is operated. However, charges are generated as the EMCCD is read out. Clock Induced Charges (CIC), a well know source of noise affecting all kinds of CCDs, were typically measured in the range of 0.1 to 0.01 electron per pixel per frame\cite{2008AIPC..984..148T, 2006SPIE.6276E..44W, techreport-minimal} (for a 512 x 512 CCD97 frame transfer EMCCD from E2V Technologies) and quickly dominate the dark current or even the photon flux as the frame rate in increased. Thus, in order to make photon counting efficient at low flux with an EMCCD, the CIC must be reduced to a minimum. Some techniques were proposed to reduce the CIC \cite{2001sccd.book.....J, 2004ASSL..300..611G, 2004SPIE.5499..219D, 2004SPIE.5499..203M, techreport-minimal, 2006sda..conf..303T} but so far, no commercially available CCD controller was able to implement all of them and get satisfying results.                                                

CCCP, a CCD Controller for Counting Photons, has been designed with the aim of reducing the CIC generated when an EMCCD is read out. It is optimized for the driving of EMCCDs at high speed, both vertically and horizontally, but may be used for driving classical CCDs as well. Using this controller, CIC levels as low as 0.001 -- 0.0018 event per pixel per \textit{frame} (as opposed to per \textit{transfer}) were measured on the 512 x 512 CCD97 EMCCD from E2V Technologies operating in \textit{inverted} mode (as opposed to \textit{non-inverted} mode). The impact of this level of CIC on the photon counting efficiency of an EMCCD will be discussed in this article. Data gathered using the controller will also be presented.

\begin{figure}[tbp]
\begin{center}
\includegraphics[width=\figurewidth]{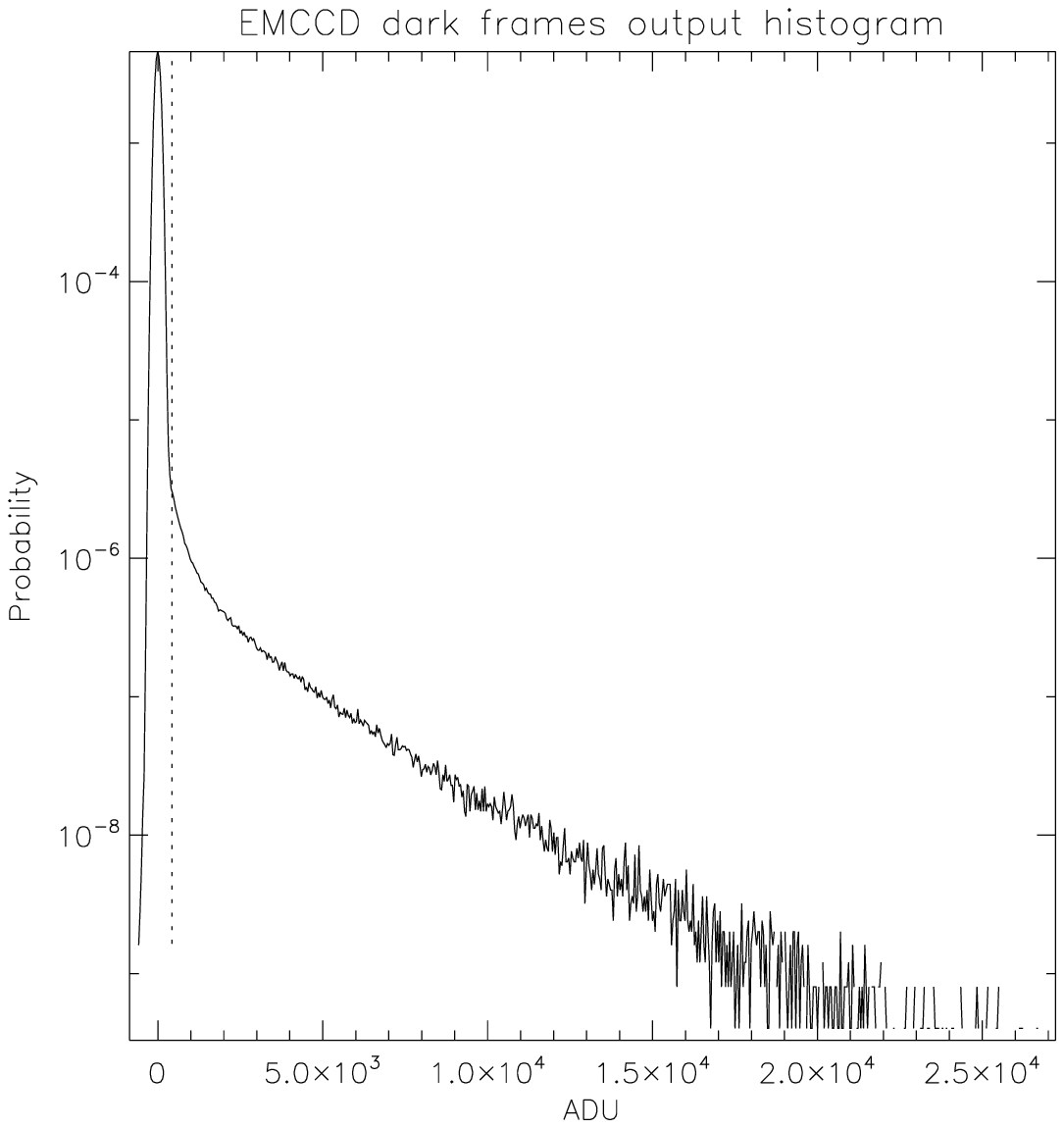}
\includegraphics[width=\figurewidth]{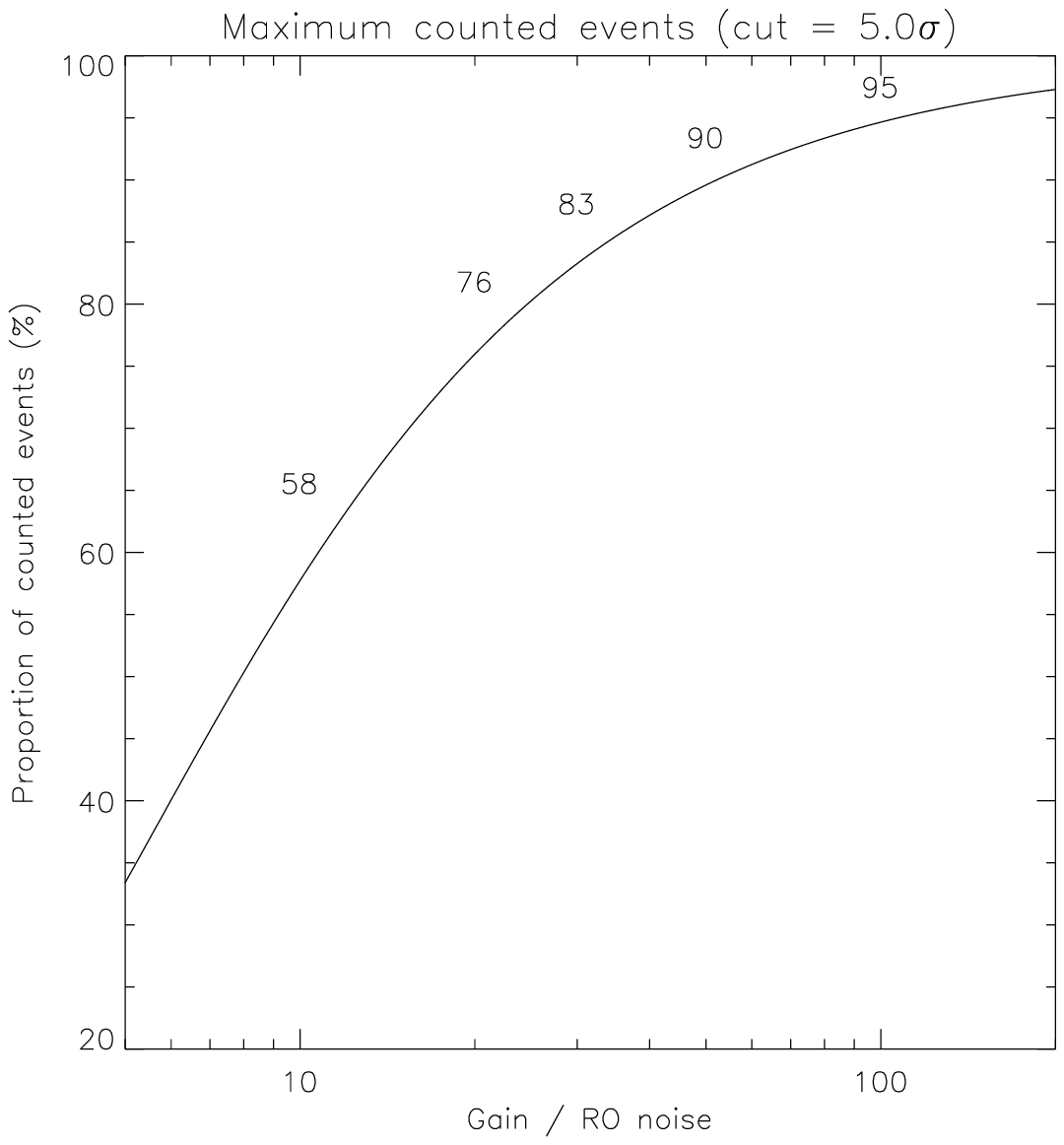}
\caption{\textbf{Left}: Histogram of an EMCCD operated under low flux, at an EM gain of $\sim$2750. Only a few pixels underwent more than one event per frame. The vertical dotted line shows the threshold at 5.5$\sigma$. The mean event rate is 0.0018 event per pixel per image. \textbf{Right}: Proportion of counted events as a function of the ratio of the EM gain over the readout noise. A cut level of 5$\sigma$ is used. Values for ratios of 10, 20, 30, 50 and 100 are printed.}
\label{fig::histogram}
\end{center}
\end{figure}

\section{PHOTON COUNTING WITH AN EMCCD}
Throughout this article, photon counting is referred to as being the process by which the output signal of the EMCCD is thresholded to a single value. Pixels having an output value higher than the threshold are considered having undergone one and only one event. Pixels having an output value lower than the threshold are considered not having undergone an event. This processing is opposed to the \textit{analogic} processing, where the output signal of the EMCCD is divided by the mean gain of its EM register to allow more than one event per pixel per frame to be considered. Operating an EMCCD in photon counting mode allows the excess noise factor to be reduced to a value of 1.

\subsection{Effect of gain and readout noise}
Sub-electron readout noise is not necessarily synonym of efficient photon counting. When one takes a look at the histogram of an EMCCD operated under low flux (left panel of figure \ref{fig::histogram}), he realizes that a significant amount of events may be hidden in the readout noise, below the threshold. The proportion $t$ of events lost due to a cut level, $cut$ (expressed in electrons), may be calculated by means of the following convolution:
\begin{equation}
\label{eqn::gainEffect}
t = \dfrac{\displaystyle\sum_{x=0}^{cut}{f(n,\lambda) \ast p(x,n,G)}}{1-f(0,\lambda)},
\end{equation}
where $f(n, \lambda)$ is the Poissonian probability of having $n$ photons during an integration period under a mean flux of $\lambda$ (in photon/pixel/frame) and $p(x,n,G)$ is the probability of having $x$ output electrons when $n$ input electrons are present at the input of the EM stage at a gain of $G$. This probability is defined by
\begin{equation}
\label{eqn::emOutputProb}
p(x,n,G) = \dfrac{x^{n-1}e^{-x/G}}{G^n(n-1)!}.
\end{equation}

Intuitively, the higher the cut level, the more events lost. Since the cut level is determined by the real readout noise, the factor to optimize will thus be the ratio of the gain over the readout noise. Right panel of figure \ref{fig::histogram} shows the relation between the ratio and the proportion of counted events. In order to count $\sim$ 90\% of the events, a ratio of 50 must be achieved. For a readout noise of 60 electrons (typical value at 10MHz of pixel rate), a gain of 3000 is necessary. However, one can not increase the gain without limit. The CIC will also increase with the gain, as discussed in section \ref{sect::cicHoriz}.

\begin{figure}[tbp]
\begin{center}
\includegraphics[width=\figurewidth]{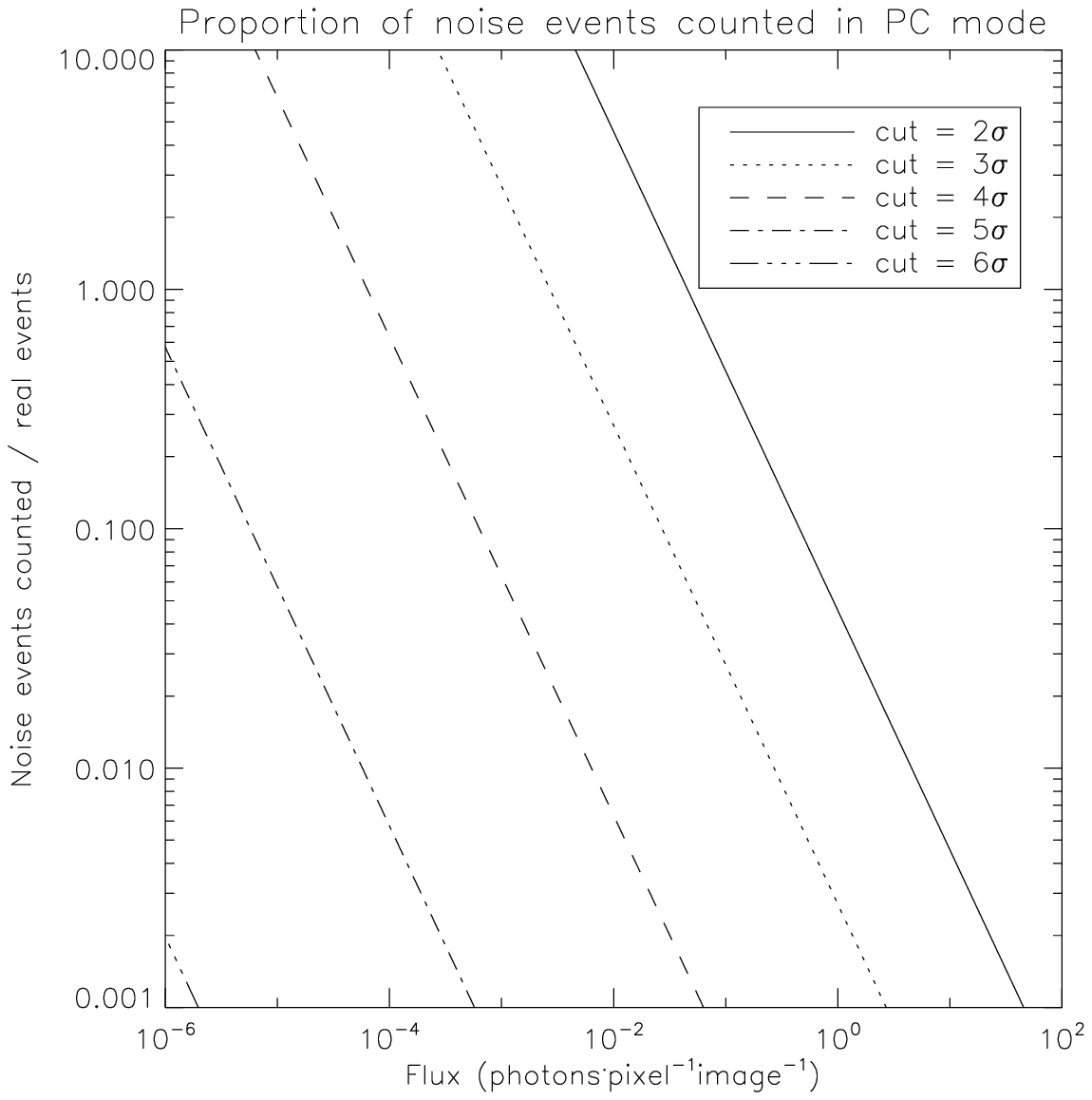}
\includegraphics[width=\figurewidth]{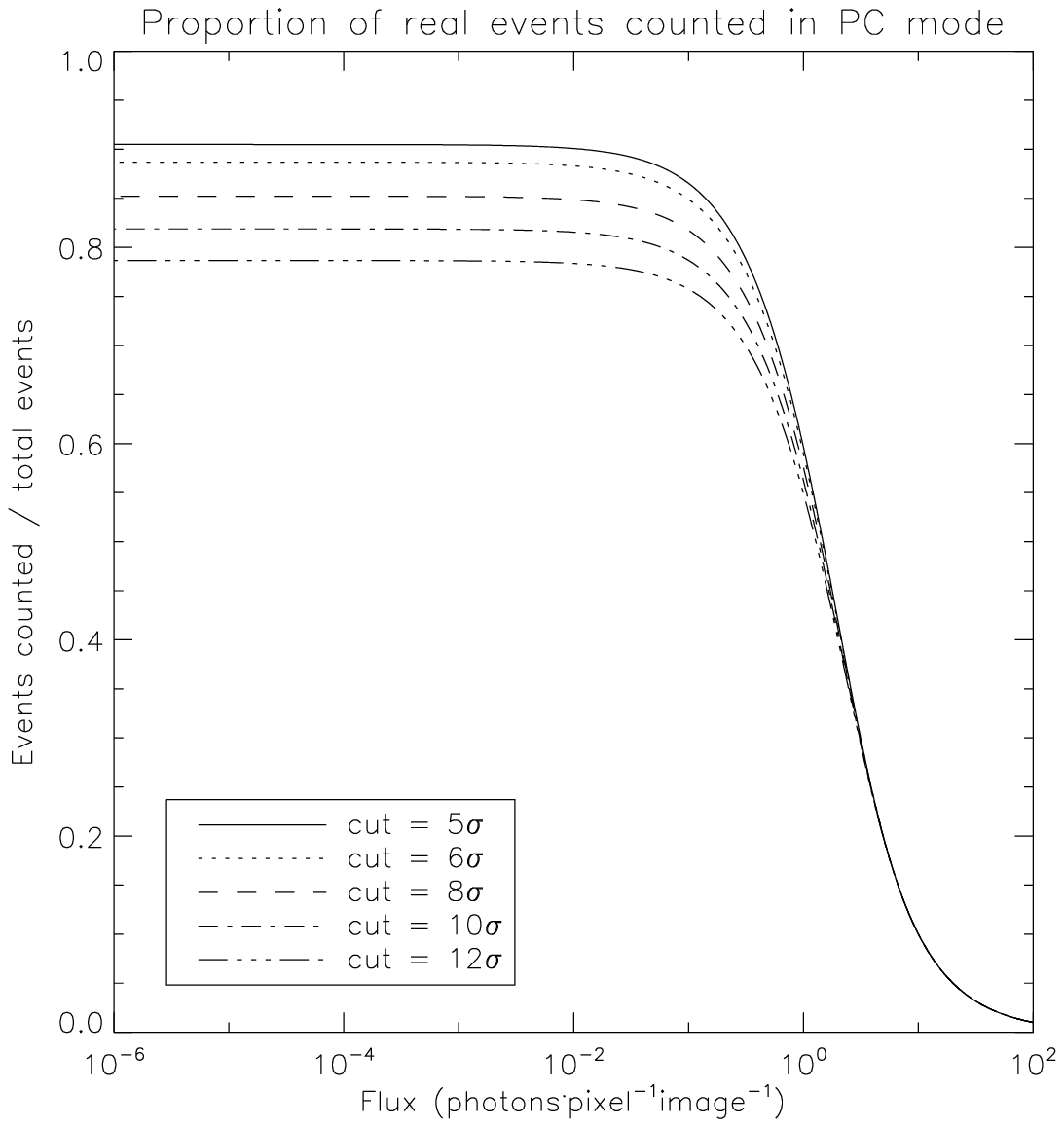}
\caption{\textbf{Left}: Proportion of noise events that are counted as real events in photon counting mode, as a function of the threshold level expressed in multiples of $\sigma$, the readout noise. \textbf{Right}: Proportion of real events that are counted in photon counting mode, for a gain/readout noise ratio of 50, as a function of the threshold level expressed in multiples of $\sigma$.}
\label{fig::overCountedEvents}
\end{center}
\end{figure}

\begin{figure}[tbp]
\begin{center}
\includegraphics[width=\figurewidth]{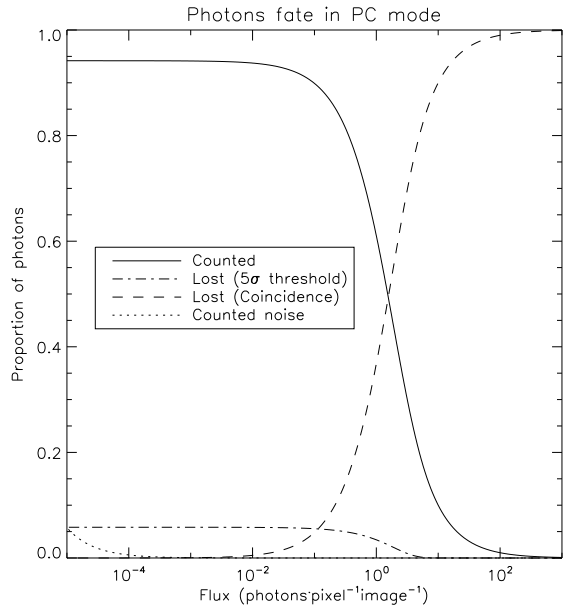}
\caption{Proportion of counted and lost events events in photon counting mode, for a gain/noise ratio of 50 and a 5$\sigma$ threshold.}
\label{fig::photonFate}
\end{center}
\end{figure}

The choice of the threshold is important and is resumed by figure \ref{fig::overCountedEvents}. Choosing a threshold that is too low yields the counting of pixels whose values above the threshold are due solely to the readout noise, as shown by the left panel of figure \ref{fig::overCountedEvents}. However, choosing a threshold that is too high yields the missing of real events (right panel). Given that the CIC level achieved with CCCP is about 0.001 event per pixel per frame, which represents the lowest count rate that will be seen in an image, a threshold of 5 $\sigma$ will cause a maximum of one pixel out of a million to have a noise event, while allowing nearly 90\% of the events to be counted (left panel of figure \ref{fig::histogram}), depending of the gain/noise ratio. Figure \ref{fig::photonFate} summarizes the fate of photons in photon counting mode.

\subsection{Coincidence losses}
The drawback of the photon counting operation is that two events occurring during a single integration time will be counted as only one. Thus, events will be lost. Poissonian statistics allows one to count $g$, the proportion of counted photons as a function of $\lambda$, the mean number of photons expected during the integration period:
\begin{equation}
\label{eqn::coincidence}
g = \frac{1-e^{-\lambda}}{\lambda}.
\end{equation}
From that equation, in order to be able to count more than 90\% of the events, the expected flux should not be higher than 0.2 event per pixel per frame.

\begin{figure}[tbp]
\begin{center}
\includegraphics[width=\figurewidth]{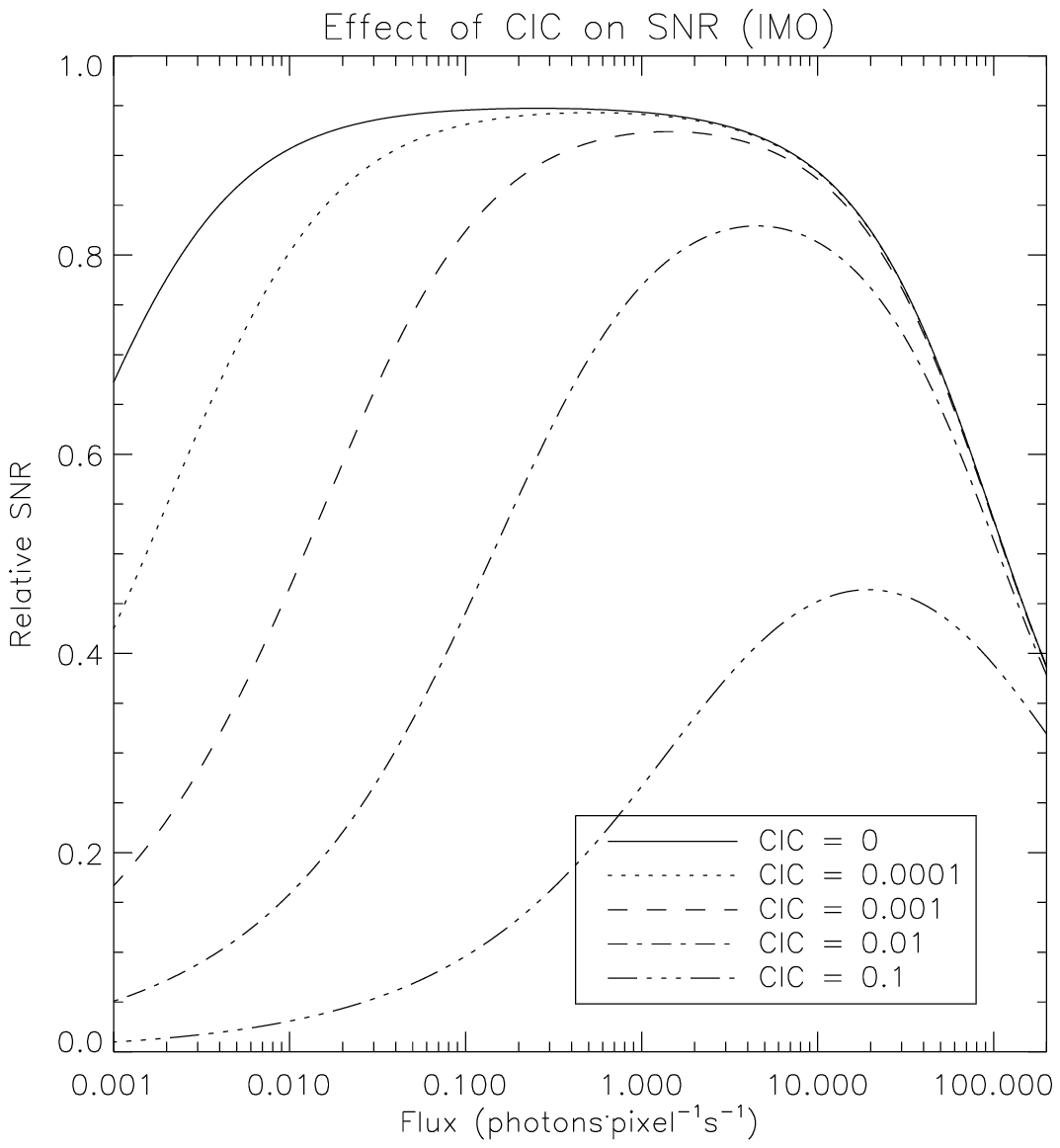}
\includegraphics[width=\figurewidth]{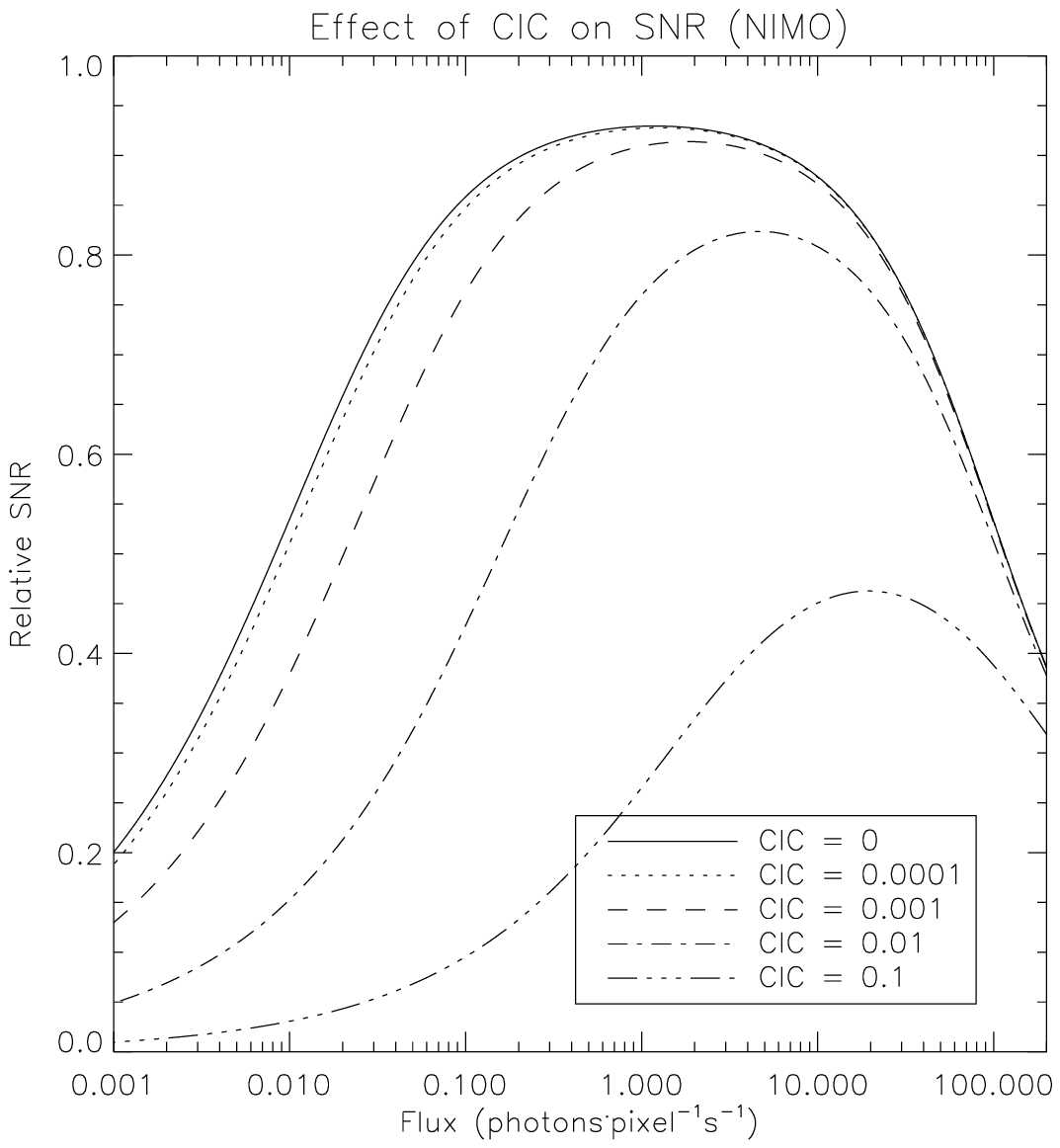}
\caption{Effect of the CIC on the SNR of an observation, compared to a perfect photon counting device having the same QE, whose noise is solely the shot noise. The simulations assume a device running at 30 frames per second, CIC is expressed as event/pixel/frame and coincidence losses are taken into account. \textbf{Left}: IMO operation: dark noise of 0.001 electron per pixel per second. \textbf{Right}: NIMO operation: dark noise of 0.02 electron per pixel per second. See text for details on the values used for the dark noise.}
\label{fig::cicComparison}
\end{center}
\end{figure}

\subsection{CIC: the dominant noise source}
Efficient photon counting with an EMCCD requires low CIC. CIC generated during the vertical transfer is dependant, among other things, of the operation mode of the CCD, namely inverted or non-inverted. As specified in Ref.~\citenum{techreport-minimal}, the amount of CIC generated during the vertical transfer could be lowered by a factor of $\sim$30 by switching from inverted to non-inverted mode, at the price of an increased dark noise. However, the surface dark current, which is suppressed by the inverted operation, is expected to dominate the bulk dark current by a factor of $\sim$200 at cryogenic temperatures. Thus, a reduction of a factor of 30 in CIC comes at the price of an increase of hundreds in dark signal.

Figure \ref{fig::cicComparison} shows how badly the CIC affects the SNR of an observation in photon counting mode. The left panel shows the simulation of a device in Inverted Mode Operation (IMO), while the right panel show the simulation of a device in Non Inverted Mode Operation (NIMO). The dark current used for the IMO plot is the one measured on a CCD97 at -85\Deg C (see temperatures considerations in section \ref{sect::temperatureCTE}) and the dark current used for the NIMO plots is calculated from equation 1 in Ref.~\citenum{techreport-minimal} for a 16$\times$16$\mu$m pixel at -85\Deg C (which is the size of the pixel of the CCD97). From these figures, IMO is clearly the operation mode to favour if one is able to achieve CIC levels in the range of 0.001 event/pixel/image. Very little gain in SNR could be achieved from driving the CCD in NIMO even if that would reduce the CIC further, since the noise would be dominated by the dark noise. It is not expected that the driving of the CCD in NIMO with CCCP would reduce the CIC since the CIC generated during the vertical transfer is very low (see section \ref{sect::resultsVertical} for more details).

\section{RESULTS}
This section presents data obtained with CCCP using a scientific grade CCD97 EMCCD from E2V Technologies operated at a pixel rate of 10MHz in IMO. Event rates presented in the figures represents not only the counted events (events above the photon counting threshold), but all the events generated during the read out process, including those buried in the readout noise. In order to achieve this, event rates are calculated by fitting the output histogram with the output probability function (equation \ref{eqn::emOutputProb}) of the EM stage. Least square minimization is then made to find the exact parameters of the output signal. This yields the EM gain and the mean event level at the same time and most importantly, this allows the exact event rate to be determined for low gain/readout noise ratios (recall figure \ref{fig::histogram}, right panel).

\begin{figure}[btp]
\begin{center}
\includegraphics[width=\figurewidth]{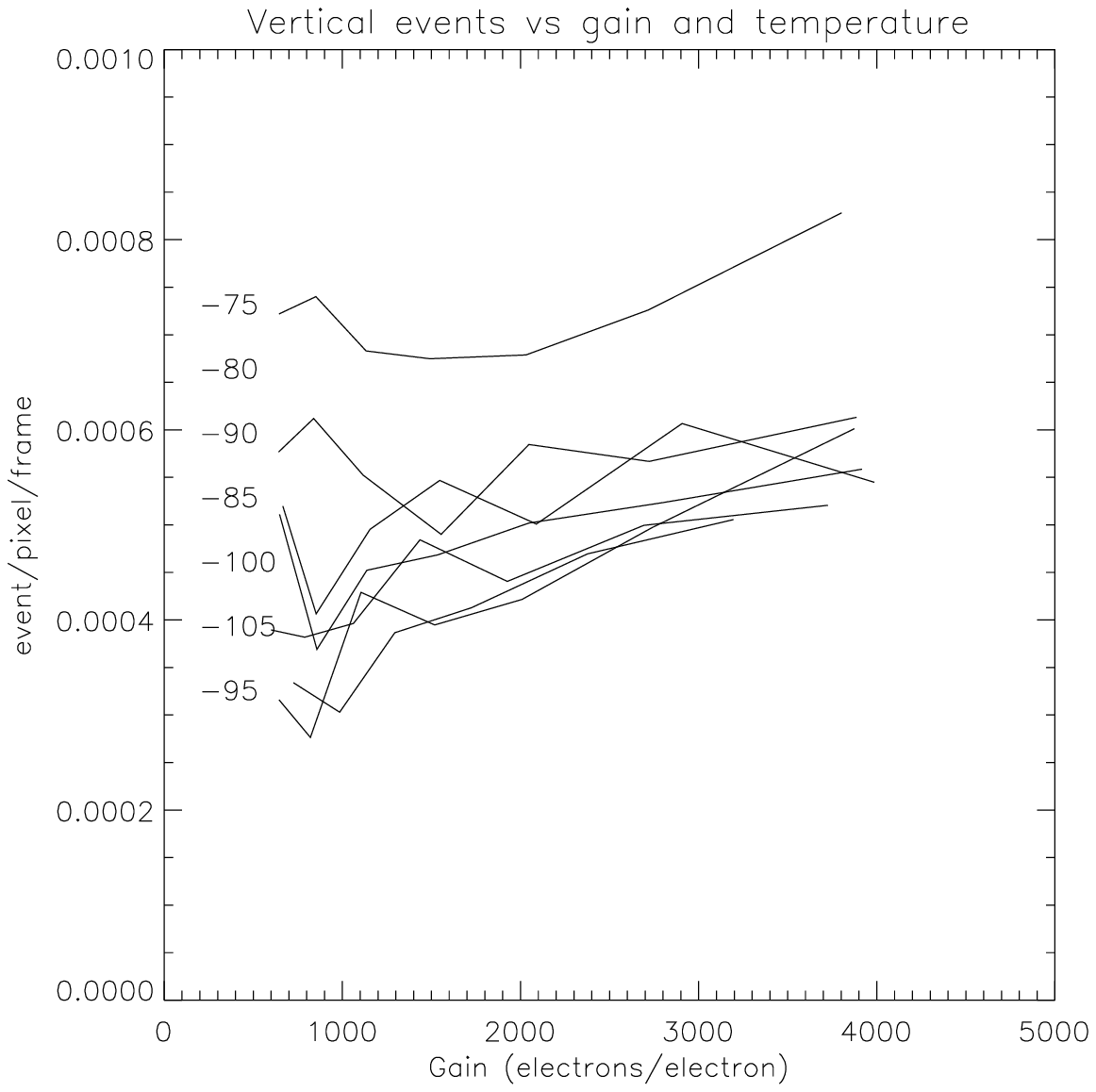}
\includegraphics[width=\figurewidth]{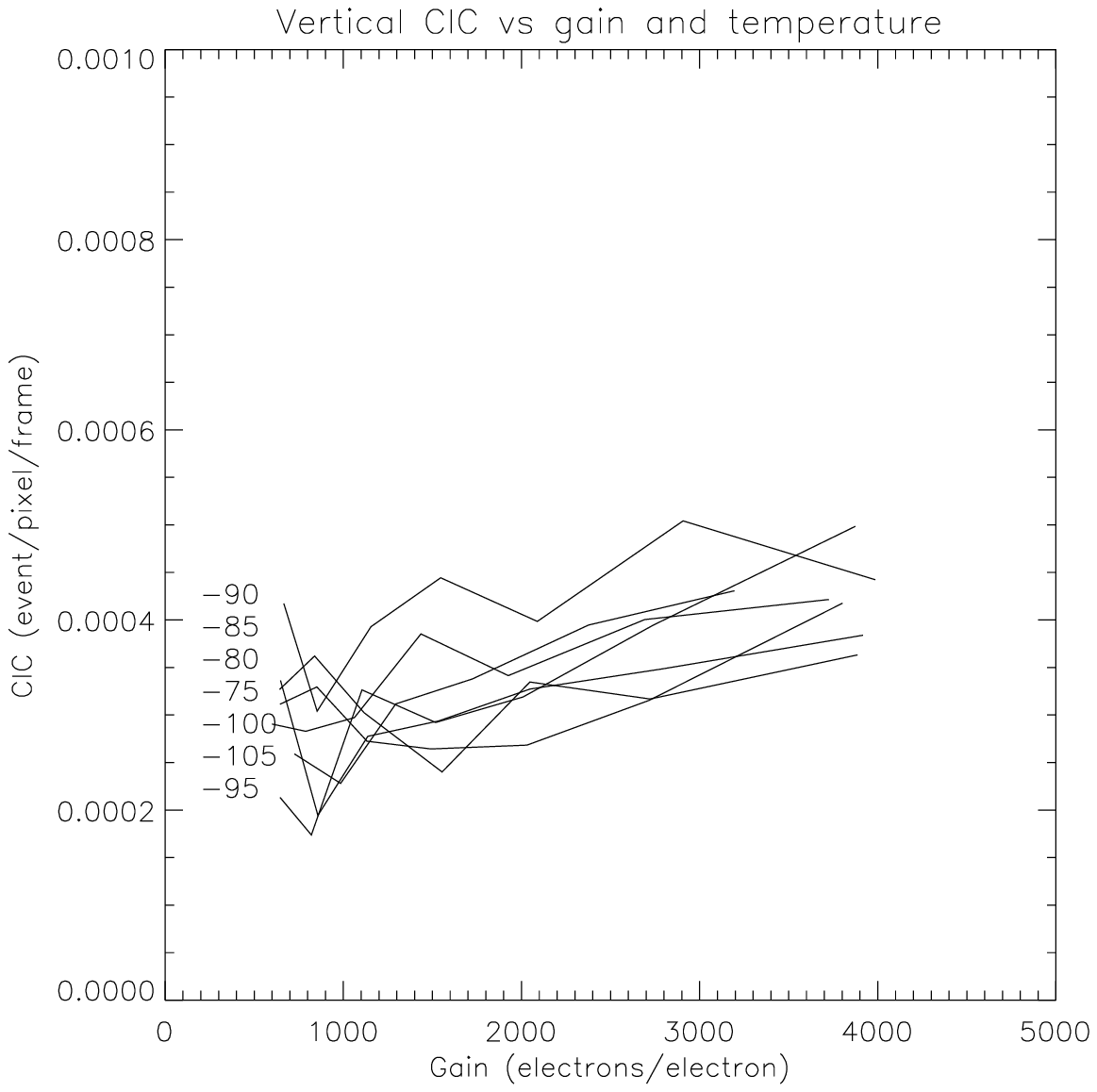}
\caption{Measurement of the events generated during the parallel transfer, for a 512 x 512 frame transfer device versus temperature and gain. The label shows the temperature, in Celcius, at which the data was acquired. Obviously, one does not expect the gain to affect the amount of vertical events. \textbf{Left}: This figure has not been corrected for the dark events that are generated during the readout process. \textbf{Right}: This figure is corrected for the dark events and should represent only the CIC events generated by the vertical transfers.}
\label{fig::vcic}
\end{center}
\end{figure}

\subsection{CIC in the vertical transfer}
\label{sect::resultsVertical}
Even when operated in Inverted mode, CCCP manages to keep the CIC low during the parallel transfer, as shown by figure \ref{fig::vcic}. The flatness of the plots is expected: the high voltage phase plays no role in the vertical transfer. The small effect of the temperature (left panel) is due to the dark noise that is generated during the read out of the CCD, as this process takes $\sim$30 milliseconds. Thus, at higher temperatures, more events are seen, even for a 0 second integration. The dark component has been suppressed from the figure in the right panel and curves in this figure should represent only the CIC generated during the vertical transfer.

\begin{figure}[tbp]
\begin{center}
\includegraphics[width=\figurewidth]{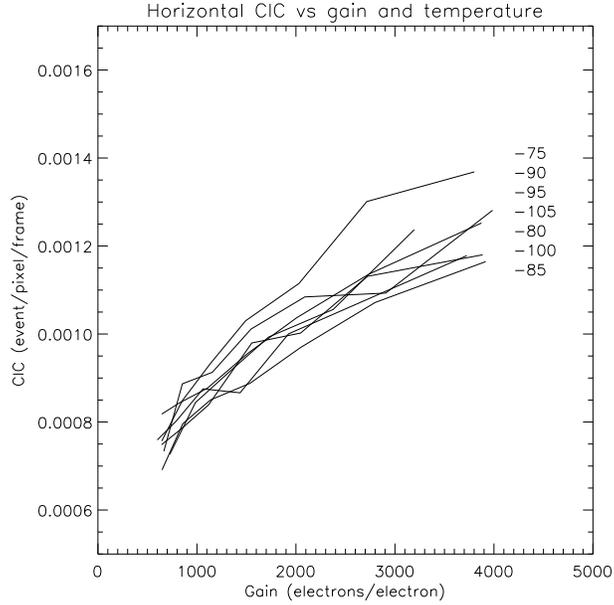}
\caption{CIC generated in the horizontal register as a function of both the EM gain and the temperature. }
\label{fig::hcic}
\end{center}
\end{figure}

\subsection{CIC in the horizontal transfer}
\label{sect::cicHoriz}
The high voltage phase of the EMCCD is meant to produce impact ionization and multiply the pixel's charge. Electrons may however be generated even in the absence of an electron at the input of the EM register. This CIC generated in the horizontal register will produce charges appearing at the output of the EM register that may be above the threshold. Dark events generated in the horizontal register will also undergo the EM amplification and appear as photon events at the output. Thus, the amount of events generated in the serial register will depend mostly on EM gain. Figure \ref{fig::hcic} shows these relations. Temperature does not play a significant role in the amount of events generated during the horizontal transfer.

\begin{figure}[tbp]
\begin{center}
\includegraphics[width=\figurewidth]{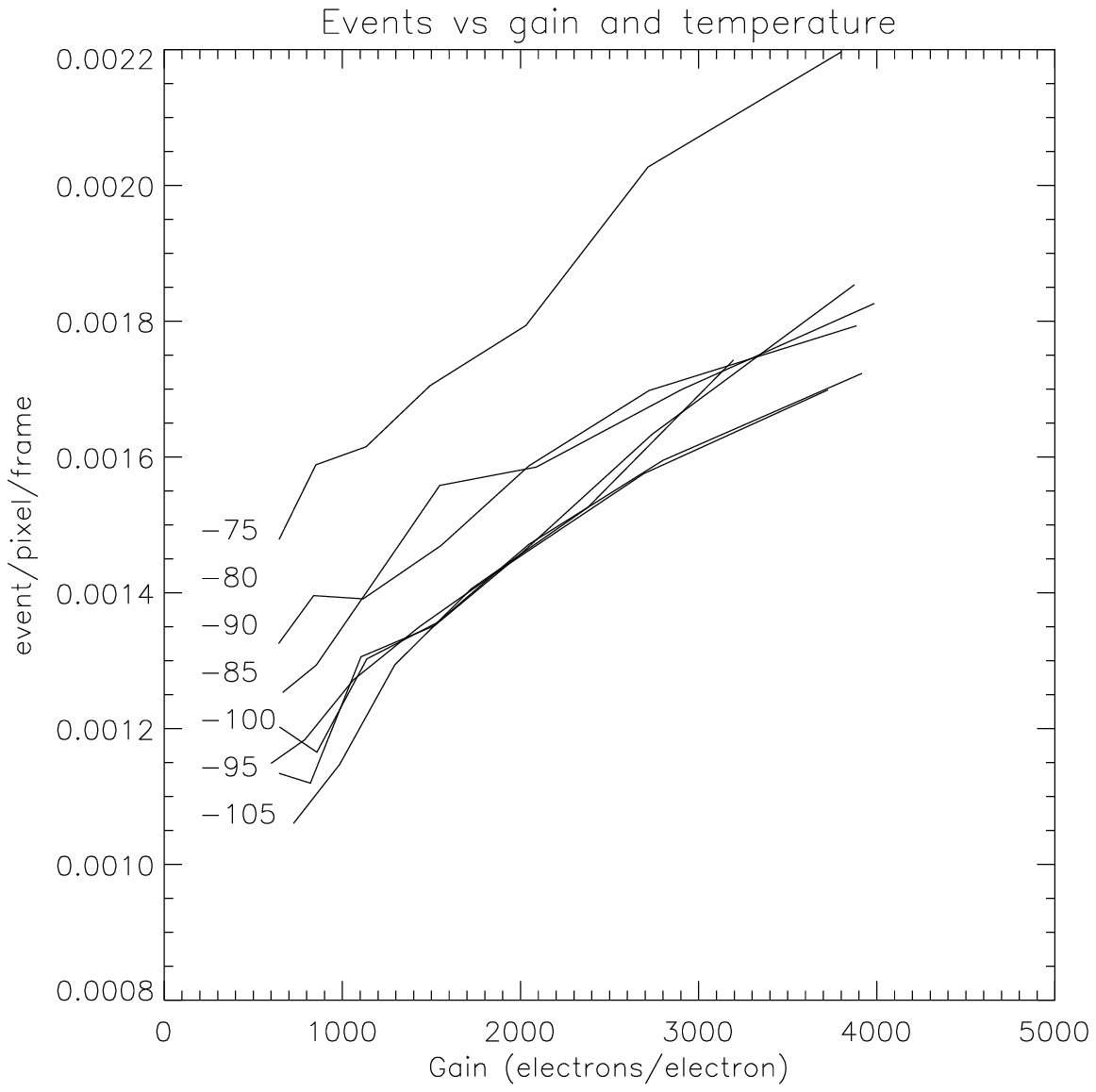}
\includegraphics[width=\figurewidth]{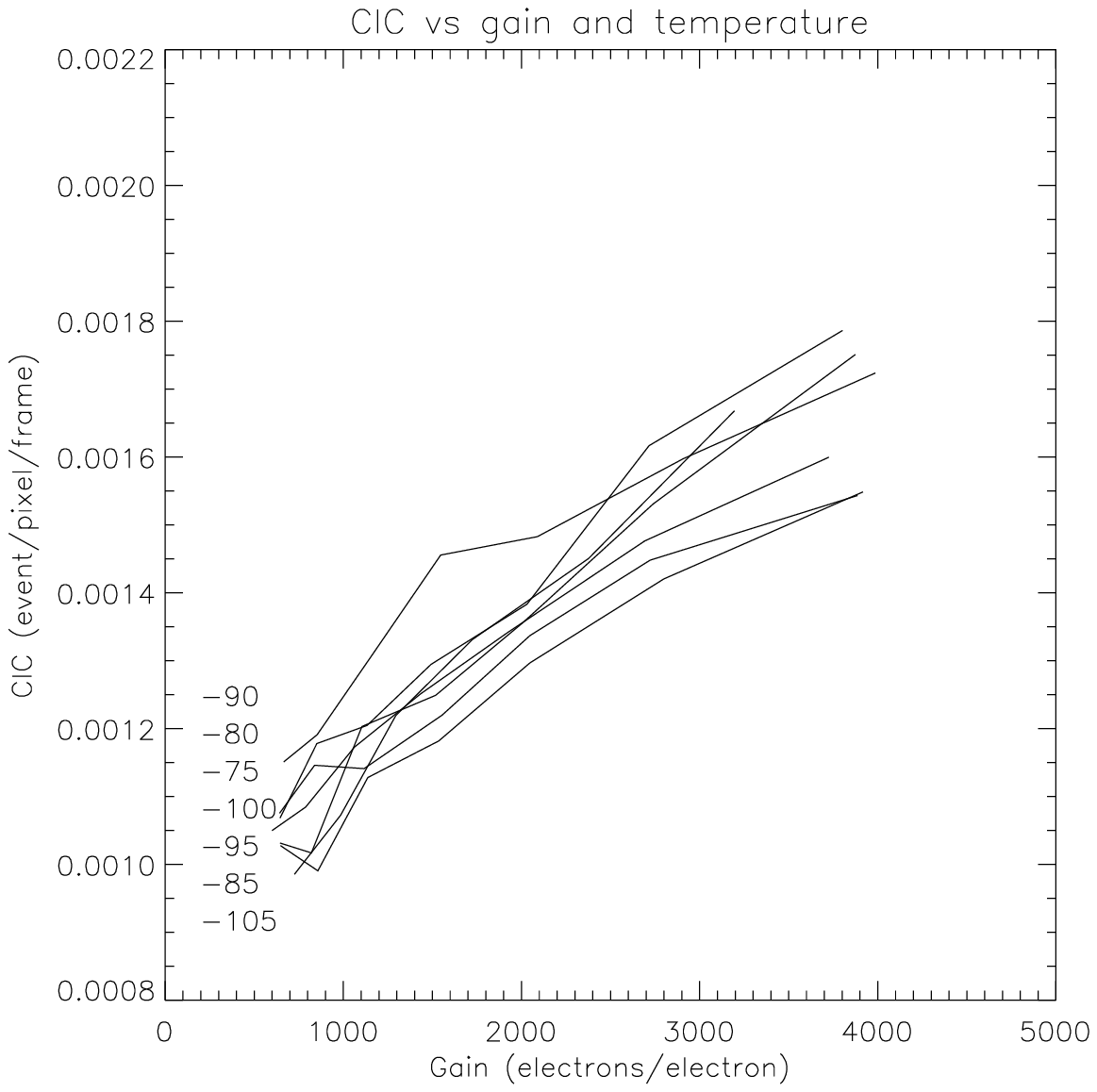}
\caption{Measurement of all the events generated during the read out process. \textbf{Left}: including dark noise. \textbf{Right}: excluding dark noise.}
\label{fig::cic}
\end{center}
\end{figure}

\subsection{Total CIC}
The total CIC is considered to be the sum of vertical and horizontal CIC. In fact, the vertical CIC presented in figure \ref{fig::vcic} is computed from the total CIC minus the horizontal CIC. Thus, the data presented at figure \ref{fig::cic} correspond to what is actually seen when the image section of the device is read. Even at EM gains as high as 4000, the total CIC measured is less than 0.002 event/pixel/frame. By comparing this figure with figures \ref{fig::vcic} and \ref{fig::hcic}, one sees immediately that the horizontal CIC is dominating over the vertical one at all gains ($>$ 800). Given its strong gain dependance, the horizontal CIC is mostly due to the high voltage clock.

\begin{figure}[tbp]
\begin{center}
\includegraphics[width=\figurewidth]{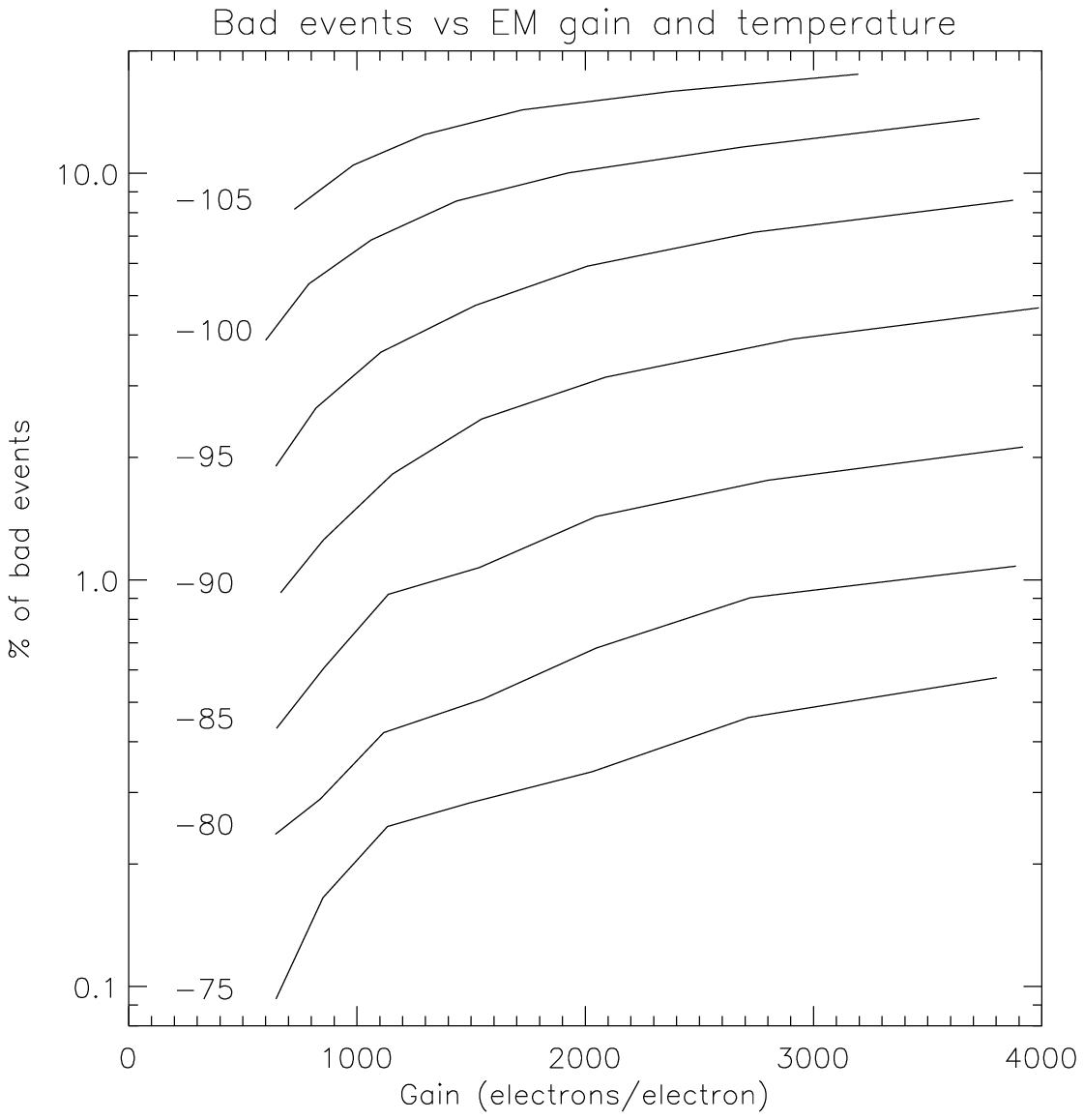}
\includegraphics[width=\figurewidth]{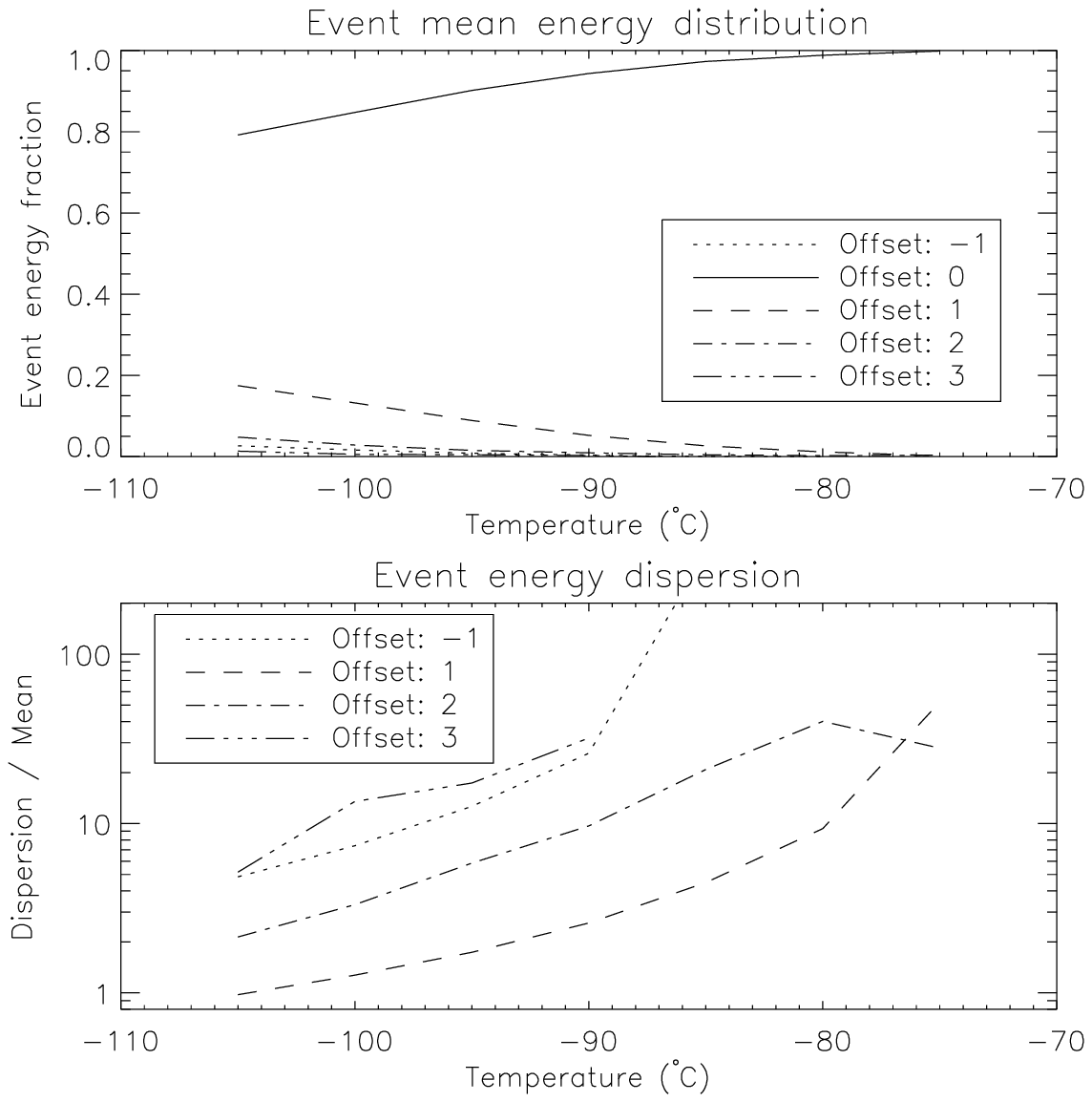}
\caption{Effect of temperature on the EMCCD. \textbf{Left}:  Fraction of \textit{bad} events (see text for a definition of bad events) seen in EMCCD images as a function of the EM gain and the operating temperature. \textbf{Right top}: Event mean energy distribution as a function of temperature. Different lines corresponds to different offset from the main (strongest) pixel of the event. Measured gain values were 3195, 2690, 2740, 2900, 2800, 2720 and 2715 for temperatures -105\Deg C, -100\Deg C, -95\Deg C, -90\Deg C, -85\Deg C, -80\Deg C and -75\Deg C, respectively. \textbf{Right bottom}: Dispersion of the energy distribution. If all events are normalized to the value of their strongest pixel, the dispersion of the proportion of the energy contained in the neighboring pixels are as plotted.}
\label{fig::bad}
\end{center}
\end{figure}

\subsection{Temperature and CTE}
\label{sect::temperatureCTE}
It is very tempting to lower the operating temperature of a CCD below -100\Deg C in order to reduce the dark noise to a minimum. However, the Charte Transfer Efficiency (CTE) in the horizontal register degrades very quickly as the temperature is lowered. The left panel of figure \ref{fig::bad} shows the amount of \textit{bad} events that are seen in an image as a function of both the gain and the temperature. Bad events are defined as being an event ($>$ 5$\sigma$) immediately followed (pixel-wise) by another. This kind of event should not occur more than once every event rate, which is only the CIC rate in this case (0.001 - 0.002 event/pixel/image). However, at temperatures below -100\Deg C, this can account for more than 10\% of the events at high gain. This is due to the events that are "leaking" into neighboring pixels as they are shifted. This phenomenon is shown by the top right panel of figure \ref{fig::bad}, where the energy distribution of the events is plotted as a function of temperatures. Each line represents the mean energy contained in a pixel at a given offset from the event (the event is at offset 0). At -105\Deg C, less than 80\% of the energy is contained in the pixel at offset 0. at -85\Deg C, this reaches 95\%.

This phenomenon would be hard to correct, since the dispersion of the energy distribution is considerable, as show by the bottom right panel of figure \ref{fig::bad}. The standard deviation of the value of the neighboring pixels around an event is at least equal to the pixel value. Thus, it would not be possible to use a post-processing based on the mean profile to compensate for the bad CTE. The shape of the profile of an event will hardly be the same as the shape of the mean profile. Thus, at high gain, temperatures below -90\Deg C should be avoided. Simulations are needed to find the perfect operating temperature, but -85\Deg C seems to be a good compromise.

\subsection{Gain stability}
\begin{figure}[tbp]
\begin{center}
\includegraphics[width=\figurewidth]{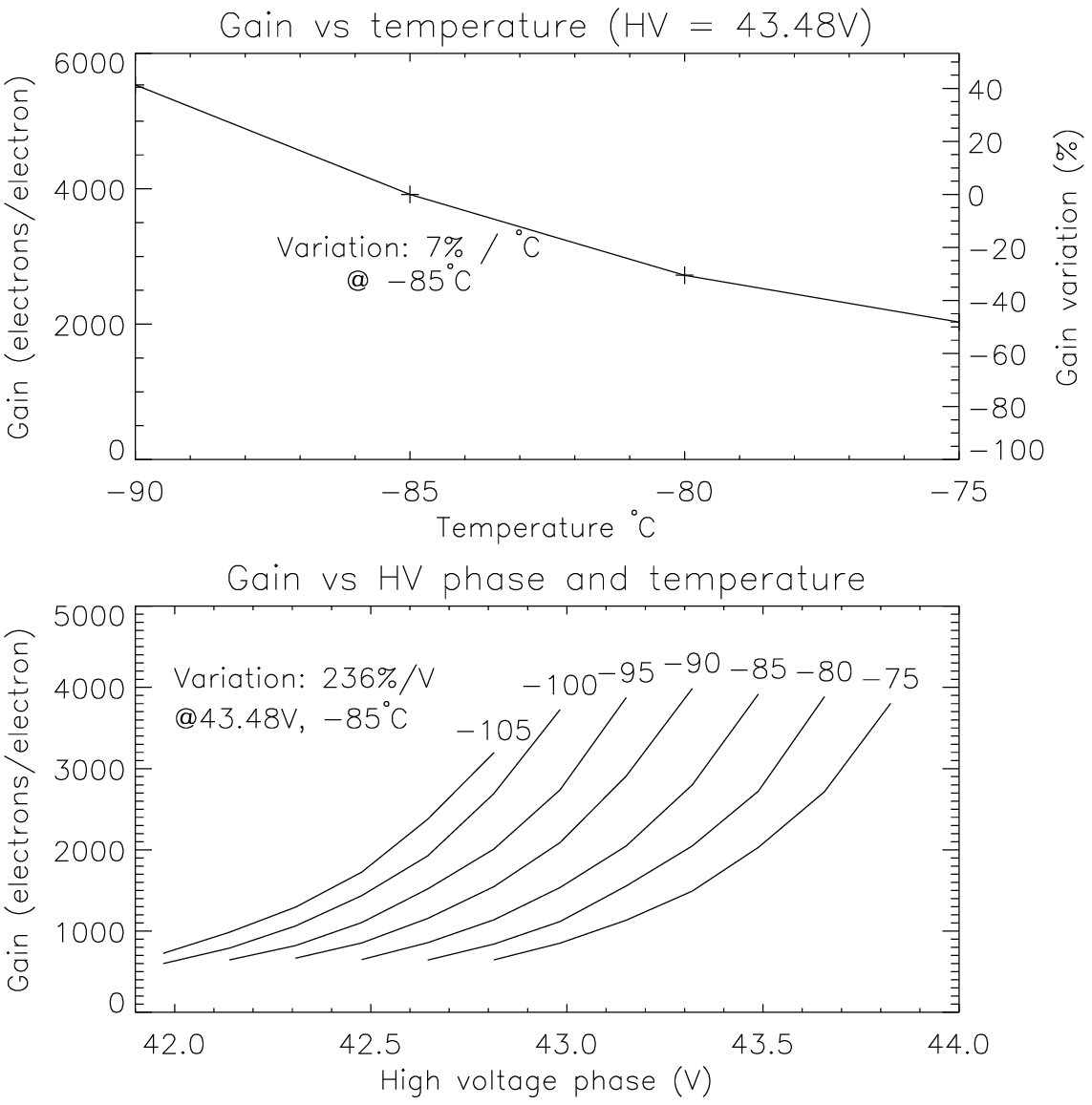}
\includegraphics[width=\figurewidth]{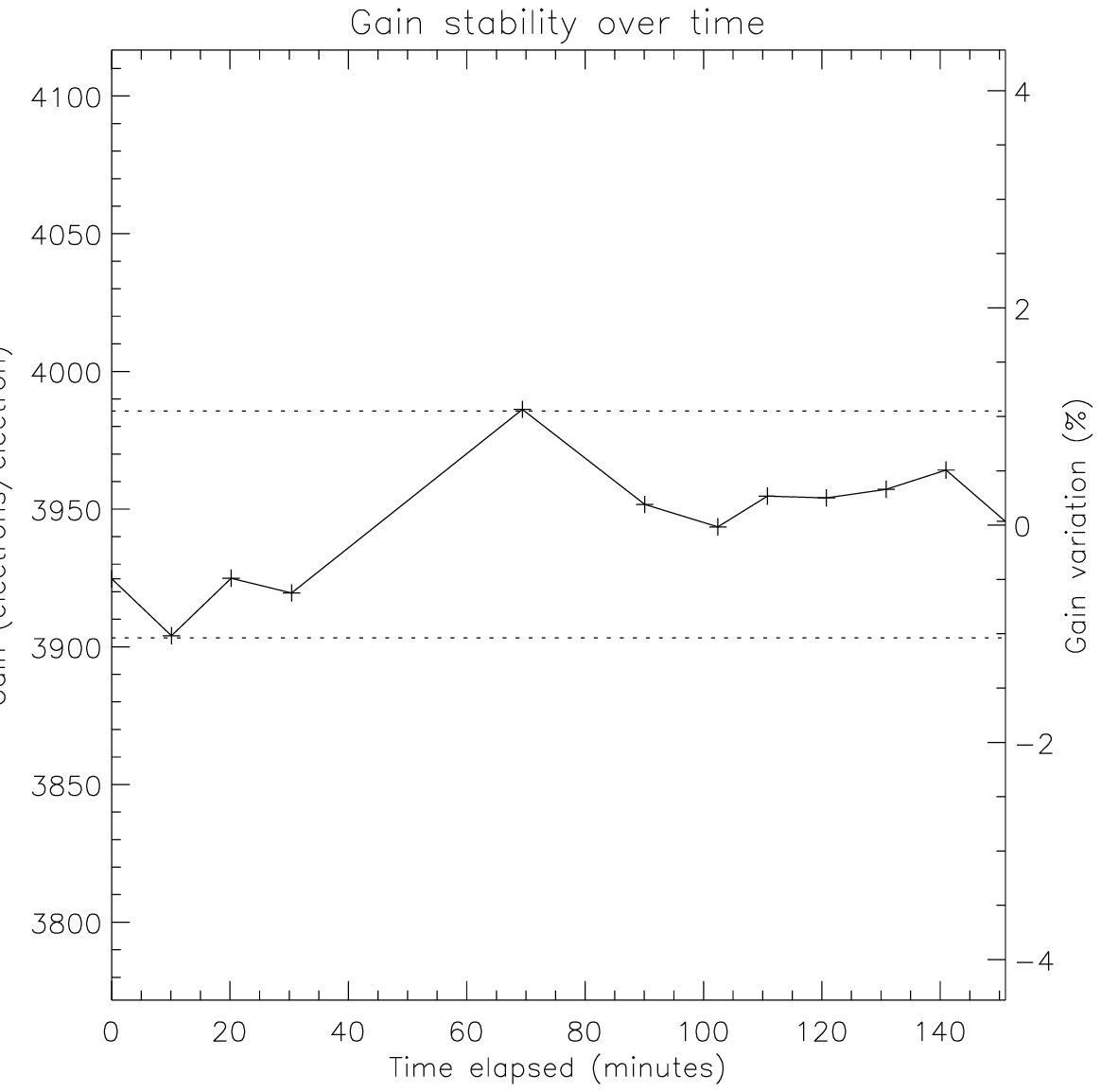}
\caption{\textbf{Left}: Effect of the temperature and the high voltage clock on the gain. At -85\Deg C and gain of $\sim$3900, the gain variation is 7\% per \Deg C and $\sim$0.2\% per mV. \textbf{Right}: Stability of the EM gain through time. The temperature variation was measured to be 0.3\Deg C peak-to-peak around -85\Deg C and the expected gain variation due to the temperature is shown by the two dotted lines.}
\label{fig::stability}
\end{center}
\end{figure}

The stability of the EM gain is not very critical for photon counting operation. However, for amplified operation, it is mandatory to have a good gain stability to ensure photometric continuity across multiple images. Two factors affects the EM gain: the high voltage phase and the temperature. From data shown in figure \ref{fig::stability}, left panel, one sees that in order to have a $\pm$1\% stability on the gain, one must have no more than $\pm$0.14\Deg C temperature variation and $\pm$5mV variation on the high voltage clock (at -85\Deg C, 43.48V HV clock, which gives a gain of $\sim$3900). However, at lower gains, the constraint on the relative gain variation per \Deg C and per mV relaxes.

The right panel of figure \ref{fig::stability} shows data acquired over time, showing a gain variation of $\sim\pm$1\%. The two dotted lines shows the variation that is expected from the temperature stability of the test dewar, which is about 0.3\Deg C peak-to-peak. From this data alone, one can not tell if the stability of the high voltage clock is sufficient; gain variation due to the high voltage clock variation could be hidden in the temperature variations. However, this data shows that precise temperature control is mandatory if less than 1\% variation on the gain is expected, at high gain. Measurements of the high voltage clock with an oscilloscope showed no more that 5mV variation through an image and over hours of operation. In fact, the oscilloscope was the limiting factor in this measurement.

\begin{figure}[tbp]
\begin{center}
\includegraphics[width=\figurewidth]{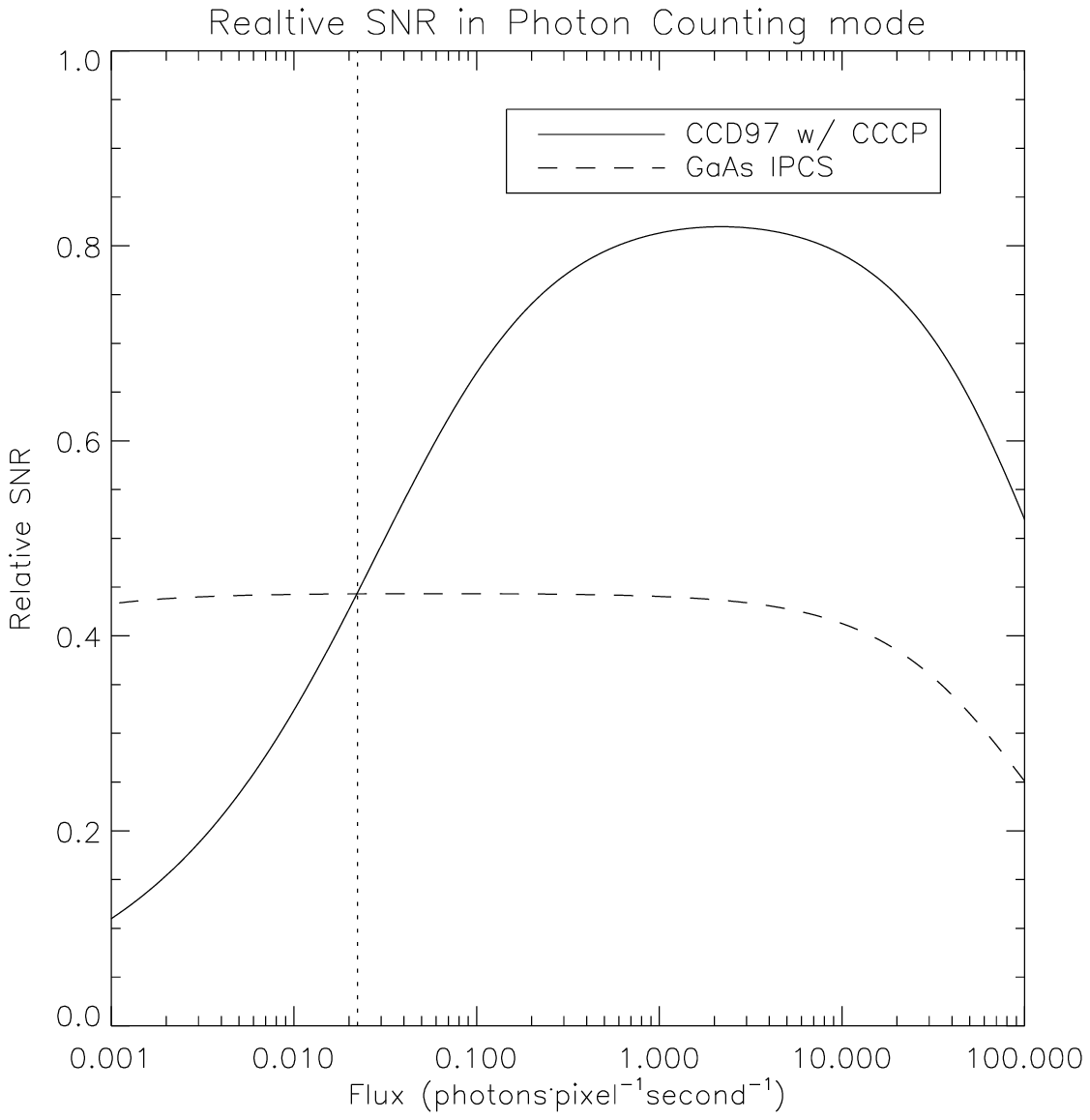}
\includegraphics[width=\figurewidth]{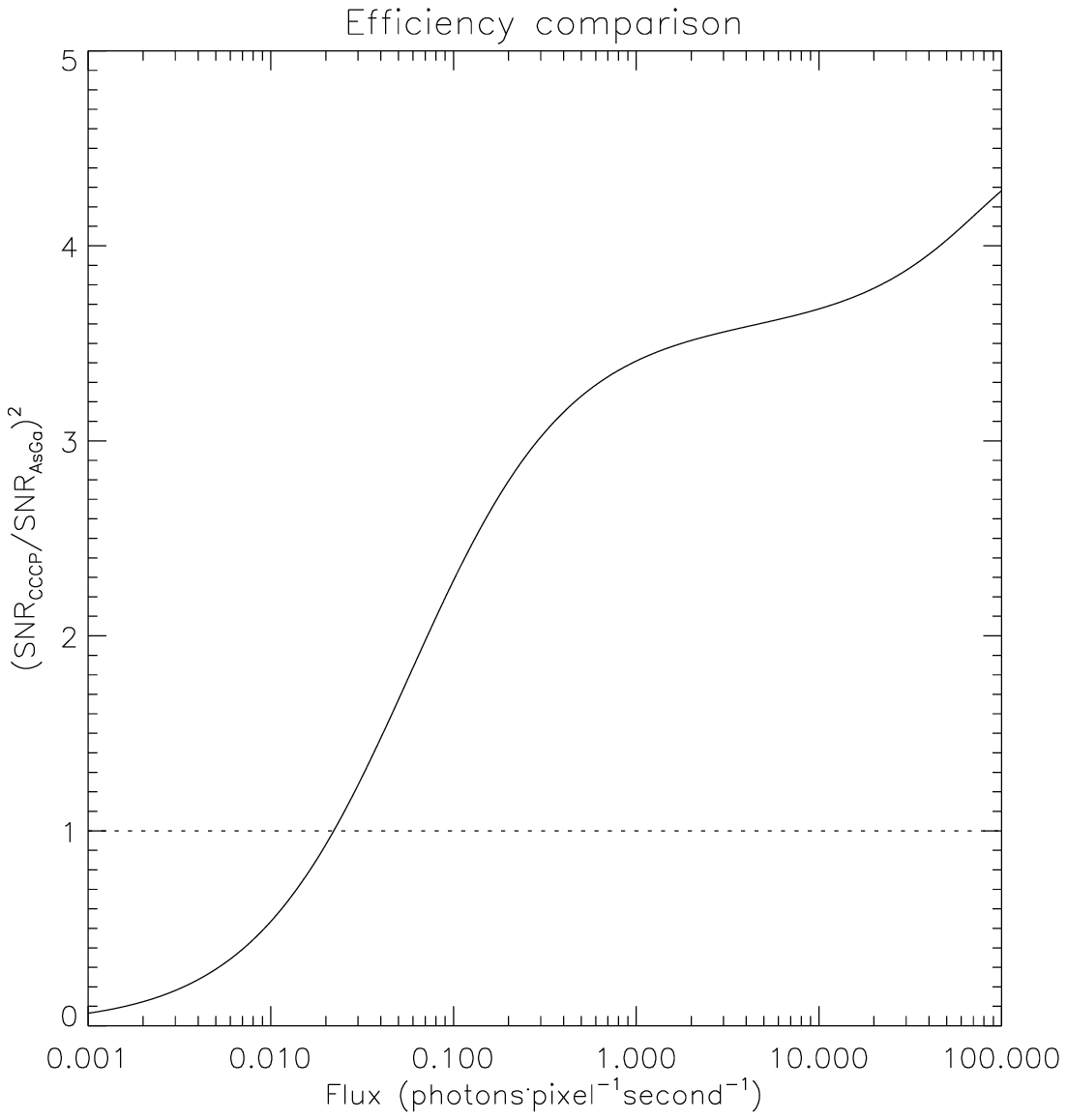}
\caption{Comparison between CCCP in photon counting mode and photocathode-based Image Photon Counting Systems (IPCS). Parameters for CCCP: gain of 3000, readout noise of 60 electrons, CIC of 0.0015 electron/pixel/frame, dark noise of 0.001 electron/pixel/second (IMO), threshold of 5$\sigma$, 30 frames per second, quantum efficiency of 80\%. Parameters for the IPCS: readout noise of 0 electron, dark noise of 0.0001 electron/pixel/second, 60 frames per second, quantum efficiency of 20\%, coincidence losses are expected to occur if two events fall in the same 3x3 pixels box. The dotted line marks the flux of equal SNR (0.022 photon/pixel/second). \textbf{Left}: SNR of both systems compared to a perfect photon counting system (QE 100\%, only shot noise). \textbf{Right}: Compared observing time efficiency of CCCP vs the GaAs IPCS. The dotted line show a relative efficiency of 1. }
\label{fig::snComparison}
\end{center}
\end{figure}

\section{CCCP efficiency}
It is of interest to compare the expected efficiency in photon counting mode of an EMCCD driven by CCCP with the photon counting systems actually in operation. In order to have zero readout noise, GaAs photocathode-based image amplifiers are often placed in front of a fast read-out CCD\cite{2002PASP..114.1043G, 2008arXiv0805.0168H}. Photons hitting the photocathode are amplified several hundreds of thousands times and produce a bright spot of a few pixels wide on the imaging CCD. Centering must then be made on these spots to recover the exact location of the incident photon. Coincidence losses on these Image Photon Counting Systems (IPCS) is thus higher than on an EMCCD in photon counting mode: two incident events that are located near from one another will be counted as only one. However, the IPCS do not suffer from CIC and their dark noise is typically an order of magnitude lower than the one of a CCD. The biggest drawback of these systems is their limited Quantum Efficiency (QE): the photocathode itself has a peak QE of $\sim$25\%. QE of EMCCDs are the same as classical CCDs and can be higher than 80\% on a wide spectral range (450-800nm) and peak at $>$95\%.

Figure \ref{fig::snComparison} compares the efficiency of such a photocathode-based system and the CCD97 driven by CCCP. The characteristics of each system is given in the caption of the figure. When taking into account all the sources of noise (dark noise, CIC, readout noise) and losses (coincidence, threshold), the left panel of the figure shows that en EMCCD will outperform an IPCS for incident fluxes higher than 0.022 photon per pixel per second (1 photon per 45 seconds). Thus, for the same pixel size, the gain in observing efficiency will be as shown in the right panel. The observing efficiency is defined as being the time it takes to reach a given SNR at a given flux. The efficiency of the EMCCD in IMO at low flux could be raised by lowering the frame rate since the noise is dominated by the CIC in this flux regime. NIMO operation of the EMCCD with CCCP would not benefit from the lowering of the frame rate as the image would be quickly dominated by dark noise (figure \ref{fig::cicComparison}).

\section{Conclusions}
Tests made with CCCP on a CCD97 EMCCD shows that the CIC can be greatly reduced without having to resort to Non Inverted Mode Operation. The low level of CIC of 0.001 -- 0.0018 event/pixel/frame, depending on the gain, allows one to use an EMCCD in photon counting mode and be more efficient than a GaAs IPCS at fluxes higher than $\sim$0.02 photon/pixel/second.

Analogic (amplified, no threshold) and conventional (no amplification) operation of the EMCCD with CCCP is also possible: the controller has the possibility to read the two outputs of the EMCCD. So far, only the CCD97 was tested, but the controller is technically able to drive other EMCCD as well as classical CCD. CCCP is designed to operate the vertical and horizontal clocks at the maximum speed specified by manufacturers, allowing fast read-out. Lower speed operation is also possible, down to a few kHz of pixel rate.

\bibliography{spie_l3_2008}
\bibliographystyle{spiebib}
\clearpage
\end{document}